# High pressure high temperature (HPHT) synthesis and magnetization of Magneto-Superconducting RuSr$_2$(LnCe$_2$)Cu$_2$O$_{12.25}$ (Ru-1232) compounds (Ln = Y and Dy)

V.P.S. Awana[*,$], and H. Kishan

National Physical Laboratory K.S. Krishnan Marg, New Delhi 110012, India

T. Kawashima and E. Takayama-Muromachi

Superconducting Materials Center (Namiki Site), National Institute for Materials Science, 1-1 Namiki, Tsukuba, Ibaraki 305-0044, Japan

C. A. Cardoso[$]

Center for Superconducity Research, University of Maryland, College Park, MD 20742-4111, USA

RuSr$_2$(LnCe$_2$)Cu$_2$O$_{12.25}$ (Ru-1232) compounds with Ln = Y and Dy being synthesized by high pressure high temperature (6GPa, 1200$^0$C) solid state synthesis route do crystallize in space group P4/mmm in near single phase form with small quantities of SrRuO$_3$ and RuSr$_2$(RE$_{1.5}$Ce$_{0.5}$)Cu$_2$O$_{10}$ (Ru-1222). Both samples exhibit magnetic transitions (T$_{mag.}$) at ~90 K with significant branching of zfc (zero-field-cooled) and fc (field-cooled) magnetization and a sharp cusp in zfc at ~ 70 K, followed by superconducting transitions at ~ 30 K. Both compounds show typical ferromagnetic hysteresis loops in magnetic moment (M) versus field (H) magnetization right upto T$_{mag.}$

i.e. < 90K. To our knowledge these are the first successfully synthesized Ru-1232 compounds in near single phase with lanthanides including Y and Dy. The results are compared with widely reported Gd/Ru-1222 and Ru-1212 ($RuSr_2GdCu_2O_8$) compounds. In particular, it seems that the Ru moments magnetic ordering temperature ($T_{mag.}$) scales with the *c*-direction distance between magnetic $RuO_6$ octahedras in Ru-1212/1222 or 1232 systems.

* Corresponding Author:  E-mail: awana@mail.nplindia.ernet.in

$ Presenting Author

## 1. INTRODUCTION

Possible co-existence of superconductivity and ferromagnetism in rutheno-cuprate viz., $RuSr_2(RE_{1.5}Ce_{0.5})Cu_2O_{10}$ (Ru-1222) and $RuSr_2GdCu_2O_8$ (Ru-1212) compounds with magnetic ordering temperature ($T_{mag.}$) being three times higher than superconducting temperature ($T_c$) has recently attracted a lot of attention [1-6], and references there in. These compounds are derived from that of $RBa_2Cu_3O_{7-\delta}$ or $CuBa_2RCu_2O_{7-\delta}$, with Cu in the charge reservoir being replaced by Ru such that the $CuO_{1-\delta}$ chain becomes a $RuO_{2-\delta}$ sheet [1-6]. $RuO_{2-\delta}$ sheets are responsible for ferromagnetism and the superconductivity resides in $CuO_2$ planes. Not only the appearance of bulk superconductivity but the nature of magnetic order is yet debated. In particular, for a meaningful interaction of magnetism and superconductivity the c-direction coupling of $RuO_{2-\delta}$ sheets ferromagnetism is important. For this, various magneto-superconducting rutheno-cuprates need to be synthesised with different lattice parameters.

In this article we report new results for first time on synthesis and magnetism of elongated c-lattice parameter compound $RuSr_2(Ln_1Ce_2)Cu_2O_{12.25}$ (Ru-1232) with Ln = Y and Dy and compare them with widely reported Ru-1222 and Ru-1212 compounds. The Ru-1232 rutheno-cuprate phase possess larger c parameter in comparison to Ru-1212 and Ru-1222 and thus warranting more distance between ferromagnetic $RuO_{2-\delta}$ sheets. In fact in the Ru-1232 structure three-layer fluorite-type blocks are inserted instead of a single oxygen-free R (= rare earth element) layer between the two $CuO_2$ planes of the Cu-1212 structure respectively [3,7]. We report that the Ru moments magnetic ordering temperature ($T_{mag.}$) scales with the c-direction distance between magnetic $RuO_6$ octahedras in Ru-1212/1222 or 1232 systems and thus warranting the c-direction coupling of ferromagnetic spins in ruthenocuprates.

## 2. EXPERIMENTAL DETAILS

Samples of composition $RuSr_2(Ln_1Ce_2)Cu_2O_{12.25}$ with Ln = Y and Dy were synthesised through a HPHT solid-state reaction route. For the HPHT synthesis, the molar ratio used were: $(RuO_2)$ + $(SrO_2)$ + $(SrCuO_2)$ + 3/4(CuO) + 1/4($CuO_{0.011}$) + 1/2($Ln_2O_3$) + 2($CeO_2$) resulting in $RuSr_2(Ln_1Ce_2)Cu_2O_{12.25}$. $CuO_{0.011}$ is pure Cu-metal, for which precise oxygen content is determined before use. The materials were mixed in an agate mortar. Later around 300 mg of the mixture was sealed in a gold capsule and allowed to react in a flat-belt-type-high-pressure apparatus at 6GPa and $1200^0C$ for 2 hours. Nearly no change was observed in the weight of synthesized samples, indicating towards their fixed nominal oxygen content. We believe the oxygen content of both the samples is close to nominal value i.e., 12.25. Determination of the oxygen content of synthesized samples is yet warranted to know the oxygen value for these samples. X-ray powder diffraction patterns were obtained by a diffractometer (Philips-PW1800) with $CuK_\alpha$ radiation. DC susceptibility data were collected by a SQUID magnetometer (Quantum Design, MPMS).

## 3. RESULTS AND DISCUSSION

Fig.1 depicts the X-ray diffraction patterns of $RuSr_2(Y_1Ce_2)Cu_2O_{12+\delta}$ (Ru-1232) with $\delta = 0.0$, 0.10 and 0.20 which could be readily indexed within tetragonal structure having space group *P*4/*mmm*. For details of the various phases of 1232 type, please see ref. 7. Also seen in Fig.1 are small quantities of $SrRuO_3$ and $RuSr_2(RE_{1.5}Ce_{0.5})Cu_2O_{10}$ (Ru-1222). Worth mentioning is the fact that both $SrRuO_3$ and Ru-1222 phases are magnetic and their presence beyond a minimum level can be marked in magnetization measurements M(T) at 150 K and 110 K respectively [3,4], this will be discussed later. Though, all the three oxygen contents gave nearly similar X-ray patterns, we decided to work with $\delta = 0.25$, with a pre-assumption that higher oxygen content could give rise to better superconductivity in terms of larger diamagnetic signals. Lattice parameters calculated are $a = b = 3.822(1)$ Å, and $c = 16.3336(4)$ Å for Ln = Y and $a = b = 3.827(3)$ Å, and $c = 16.3406(7)$ Å for Ln = Dy samples of the $RuSr_2(Ln_1Ce_2)Cu_2O_{12.25}$ series. The trend of their cell volumes is in line with the rare earths ionic sizes.

Figure 2 shows both zero-field-cooled (zfc) and (field-cooled) fc magnetic susceptibility versus temperature ($\chi$ vs. T) plots for the Y/Ru-1232 sample, in external fields of 10 Oe. As seen from this figure the fc magnetization curve shows an increase near 90 K ($T_{mag.}$). Furtrher, the zfc branch shows a cusp like down turn in magnetization at around 70 K. In general the magnetization behaviour of the compound can be assigned to a weak ferromagnetic transition at around 90 K. The zfc branch also shows a step like structure at around 45 K ($T_c$) and a diamagnetic transition around 35 K ($T_d$).

Figure 3 shows both zero-field-cooled (zfc) and fc magnetic susceptibility versus temperature ($\chi$ vs. T) plots for the Dy/Ru-1232 sample. The general behaviour is similar to that as for Y/Ru-1232. The fc transition is seen at 90 K. The zfc cusp is at around 70 K, the step like structure ($T_c$) at around 40 K and diamagnetic transition ($T_d$) is seen at around 35 K for Dy/Ru-1232 compound. The exact nature of Ru moments ordering in Ru-1212 [2], Ru-1222 [1] and presently studied Ru-1232 is not yet clearly understood. The magnetic ordering temperature ($T_{mag.}$) of Ru moments in Ru-1212/1222 and 1232 is respectively 140 K (2), above 110 K (3) and 90 K (present study) respectively. Interestingly the c-lattice parameters of Ru-1212, 1222, and 1232 are nearly 11.5 Å, 14 x 2 Å and 16.5 Å. It seems the Ru moments ordering temperature ($T_{mag.}$) scales with c-lattice parameter. Higher is the c- lattice partameter (effective distance between $RuO_2$ sheets) the lower is $T_{mag.}$, please see inset in Fig. 3. This concludes that Ru moments ordering in these systems is all over the sample and Ru moments interact magnetically not only in a-b but also in c-direction.

Also it is clear from χ vs. T results being shown in Fig.2 and Fig.3 that though both Y/Ru-1232 and Dy/Ru-1232 contain $SrRuO_3$ and Ru-1222 as impurity phases (XRD results), the respective $T_{mag.}$ of these foreign phases are abscent. This concludes that presently studied Y/Ru-1232 and Dy/Ru-1232 compounds are nearly single phasic in nature.

Superconductivity is seen in terms of diamagnetic transition at below $T_d$. It is known earlier that due to internal magnetic field, these compounds are in a spontaneous vortex phase (SVP) even in zero external field [4]. For $T_d < T < T_c$ the compound remains in the mixed state. Hence though the superconductivity is achieved at relatively higher temperatures, the diamagnetic response is seen at much lower temperature and that too in quite small applied magnetic fields.

To further elucidate on the magnetization of these compounds, the isothermal magnetization as a function of magnetic field at 5, 20, 50, 80 and 120 K with applied fields; -70000 Oe ≤ H ≤ 70000 Oe for Ln = Dy sample is shown in Fig.4. The saturation of the isothermal moment appears to occur above say 4 Tesla applied field at 5 K. Further increase in magnetization above say 4 Tesla is due to the contribution from Dy moments. At higher temperatures of 20, 50, 80, 100 and 120 K the near saturation of M vs. H is not seen. The presence of the ferromagnetic component is confirmed by hysteresis loop being observed at 5 K in the M vs. H plots (-1000 Oe ≤ H ≤ 1000), (see inset Fig. 4). Ru spins order magnetically above say 90 K with a ferromagnetic component within at 5 K. As far the value of higher field (> 4 T) saturation moment is concerned, one cannot extract without ambiguity the value for Ru contribution, because the contribution from Cu cannot be ignored. In an under-doped HTSC compound Cu contributes an unknown paramagnetic signal to the system [8].

Fig.5 shows the isothermal magnetization (*M*) vs. applied field (H) behaviour for Dy/Ru-1232 in low fields of -1000 Oe ≤ H ≤ 1000. Clear M vs. H loops are seen at 5, 20, and 50 K, but not at 80 K. The returning moment ($M_{rem}$) i.e. the value of magnetization at zero returning field and the coercive field ($H_c$) i.e. the value of applied returning field to get zero magnetization, are clearly seen up to 50 K. Worth mentioning is the fact that Dy (magnetic rare earth) in the compound must order magnetically below 0.5 K and Ce is known to be in tetravalent non-magnetic state hence the $M_{rem}$ and $H_c$ arising from the ferromagnetic hysteresis loops do belong to Ru only. Hysteresis loops are not seen for M vs. H plots at or above 80 K. For various hysteresis loops being observed from M vs. H plots below 80 K, the values of both $M_{rem}$ and $H_c$ decrease with increase in T. Both $M_{rem}$ and $H_c$ being observed for Ru-1232 are much higher than the reported values for other magneto-superconductor Ru-1212 and comparative to Ru-1222. For Ru-1212 the hysteresis loops are reported quite narrow [2,3,6]. This indicates that in Ru-1232 the ferromagnetic domains are less anisotropic and more rigid.

## 4. CONCLUSION

In summary, near single phase compounds of the composition $RuSr_2(LnCe_2)Cu_2O_{12.25}$ with Ln = Y and Dy were synthesised successfully through a HPHT solid-state reaction route for the first time and their brief magnetization data is reported and discussed. Ln/Ru-1232 compounds being reported here with non-magnetic Ln like Y could help in better understanding of the magnetic structure of these compounds. Neutron scattering experiments along with various other physical property experiments on such samples are warranted.

## FIGURE CAPTIONS

Figure 1. Room temperature X-ray diffraction pattern for $RuSr_2(Y_1Ce_2)Cu_2O_{12+\delta}$ with $\delta$ = 0.0, 0.10 and 0.20.

Figure 2. Magnetic susceptibility versus temperature ($\chi$ vs. $T$) plots for $RuSr_2(Y_1Ce_2)Cu_2O_{12.25}$ (Y/Ru-1232) sample.

Figure 3. Figure 2. Magnetic susceptibility versus temperature ($\chi$ vs. $T$) plots for $RuSr_2(Dy_1Ce_2)Cu_2O_{12.25}$ (Dy/Ru-1232) sample.

Figure 4. M vs.H at 5, 20, 50, 80 and 120 K with applied fields; -70000 Oe $\leq$ H $\leq$ 70000 Oe for $RuSr_2(Dy_1Ce_2)Cu_2O_{12.25}$ (Dy/Ru-1232) sample, inset shows the zoomed part of the same at 5 K with -1000 Oe $\leq$ H $\leq$ 1000 Oe

Figure 5. *M* vs. *H* plot for the Dy/Ru-1232 compound at T = 5, 20, 50 and 80 K, the applied field H are in the range of -1000Oe $\leq$ H $\leq$ 1000 Oe.

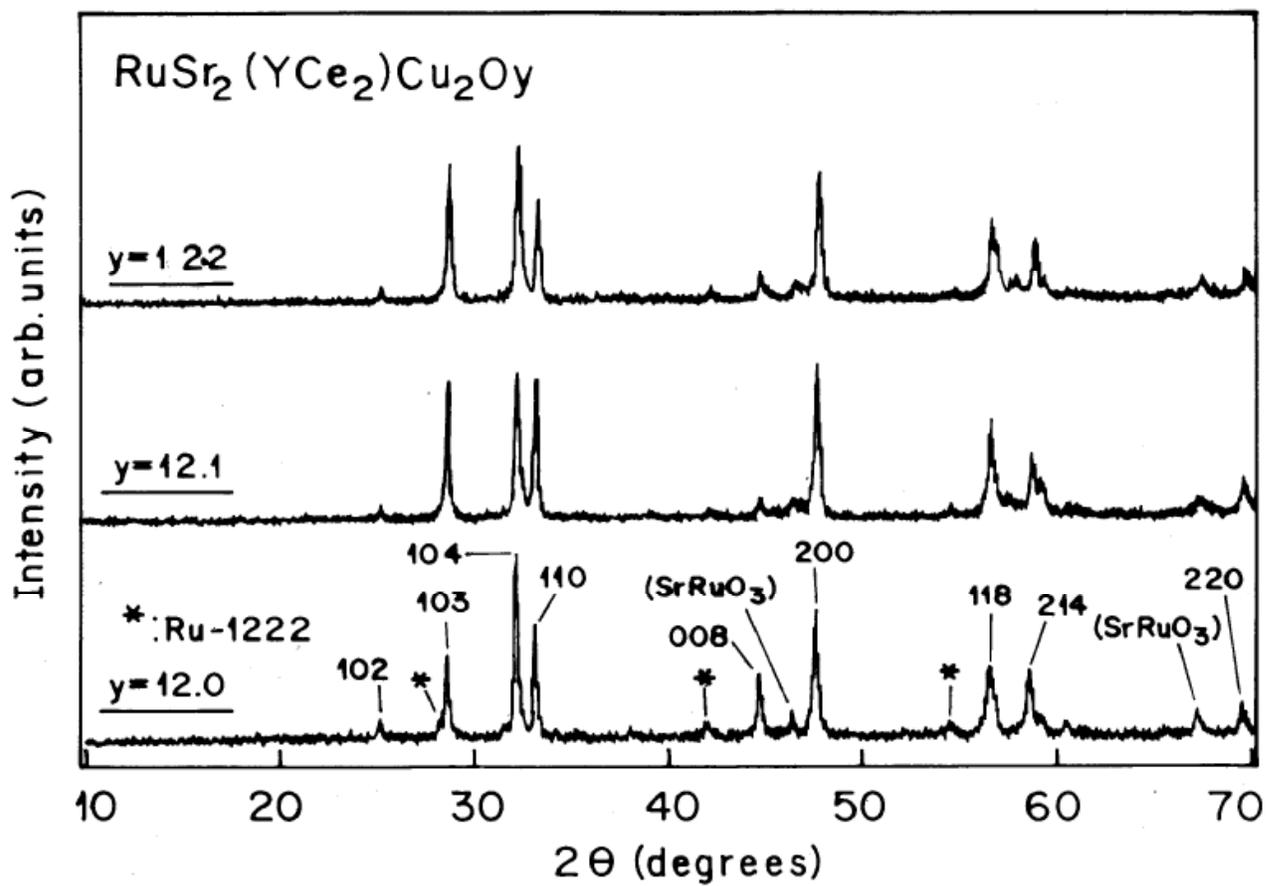

Figure 1. HU-06 (Awana et al.)

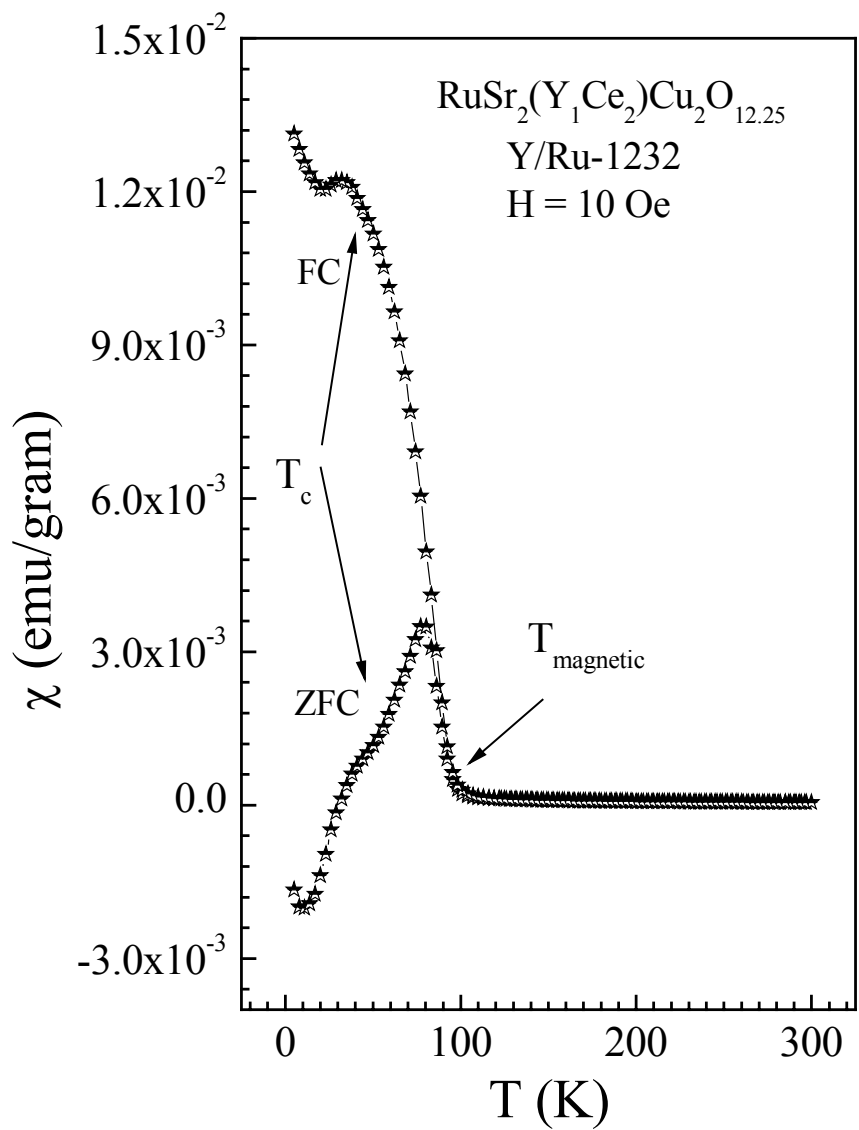

Figure 2. HU-06 (Awana et al)

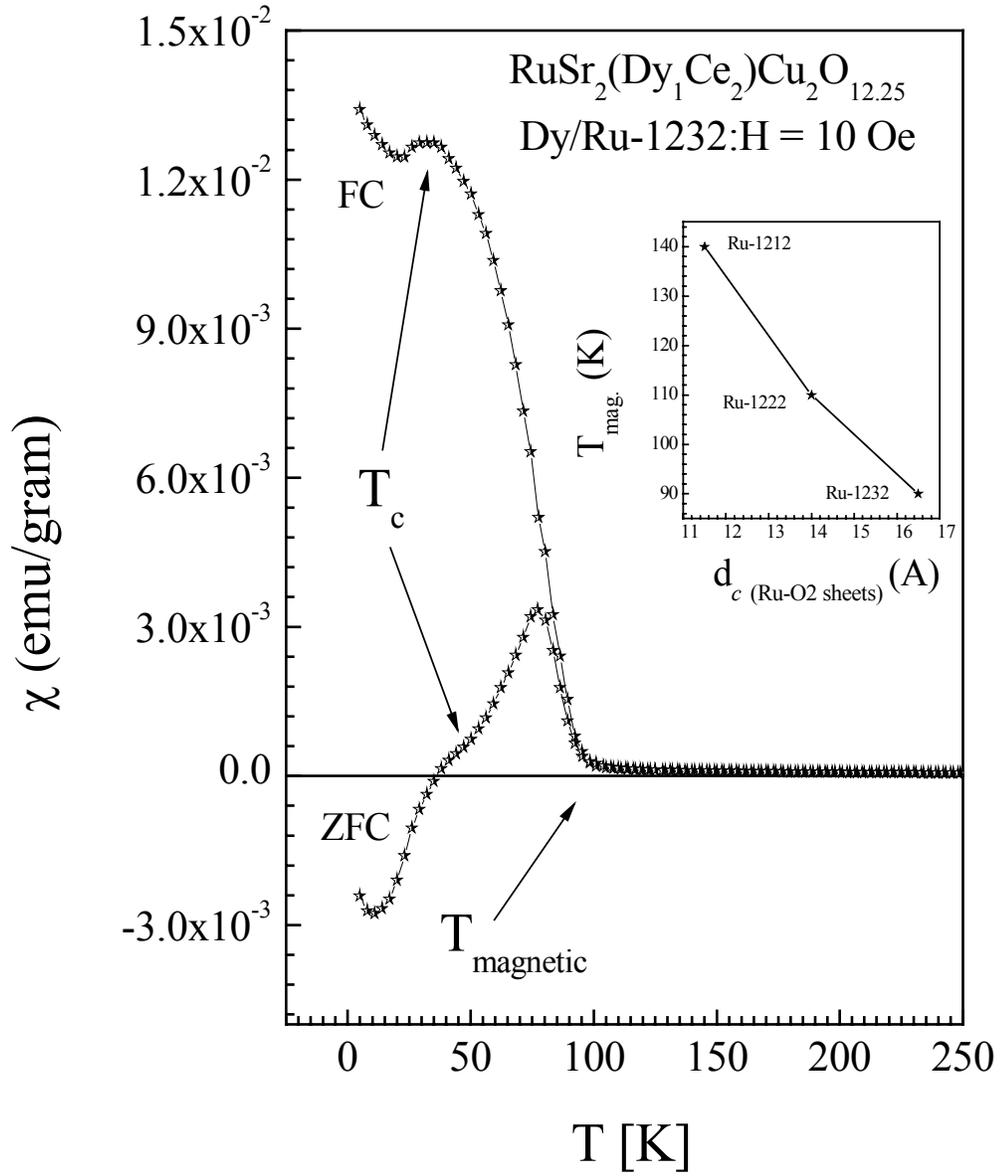

Figure 3. HU-06(Awana et al)

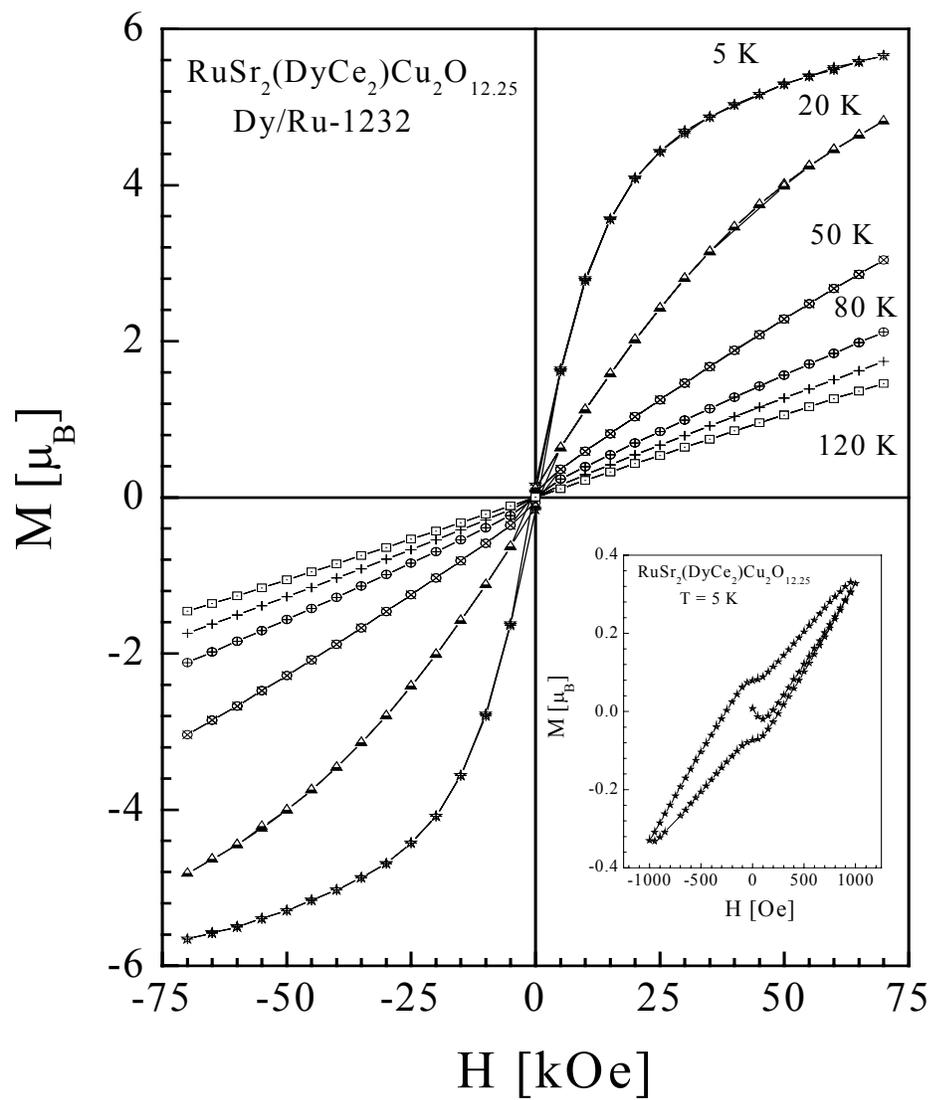

Figure 4. HU-06 (Awana et al)

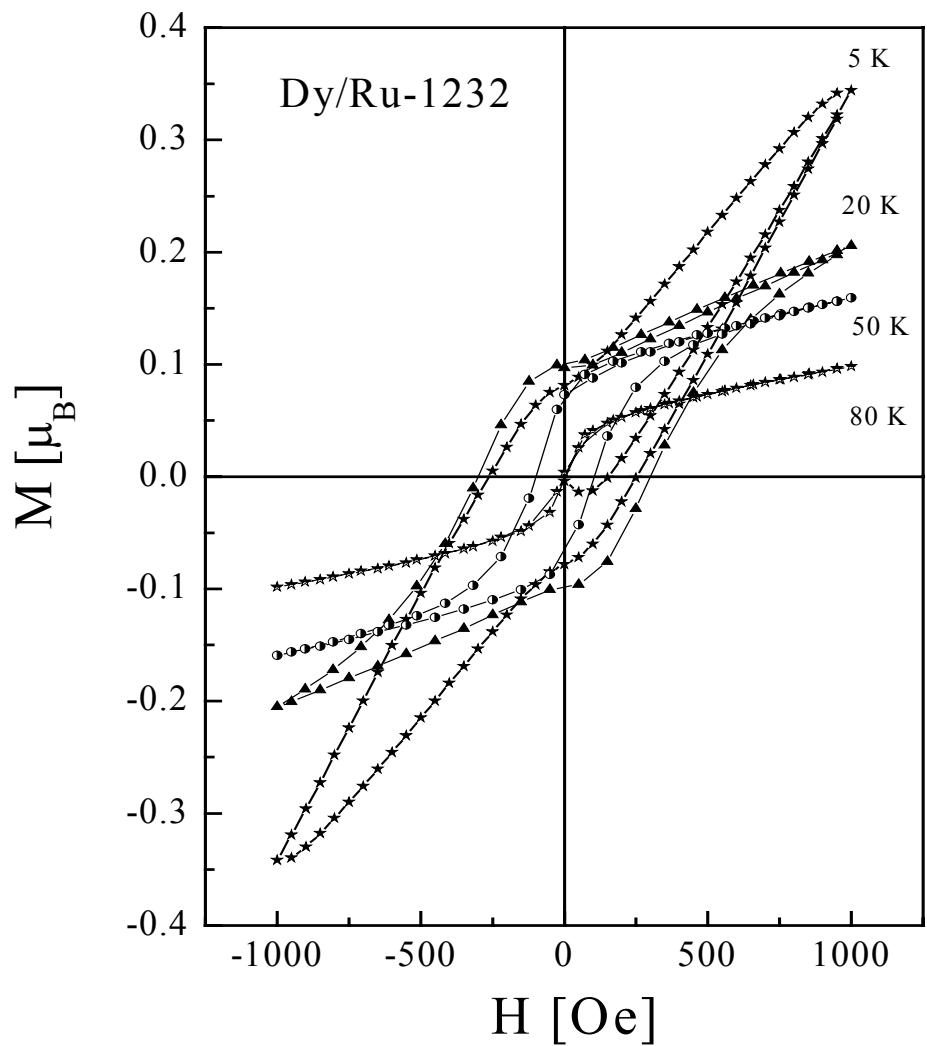

Figure 5. HU-06 (Awana et al)